\documentclass[aps,pre,twocolumn,amsmath,amssymb,superscriptaddress,showpacs,longbibliography]{revtex4-1}

\usepackage{graphicx}
\usepackage{dcolumn}
\usepackage{bm}
\usepackage{mathrsfs}

\usepackage{natbib}
\usepackage[text={7in,9.5in},centering]{geometry}

\usepackage{hyperref}
\usepackage{url}

\usepackage{epsfig}
\usepackage{amsmath}
\usepackage{amssymb}
\usepackage{textcomp}

\usepackage{color}
\usepackage{float}

\usepackage{stackengine}
\usepackage{verbatim}

\begin{document}

\title{Observation of the X17 anomaly in the $^7$Li($p$,$e^+e^-$)$^8$Be
 direct  proton-capture reaction}

\author{N.J. Sas}
\affiliation{University of Debrecen, 4010 Debrecen, PO Box 105, Hungary and
Institute of Nuclear Research  (ATOMKI), P.O. Box 51, H-4001 Debrecen, Hungary}
\author{A.J. Krasznahorkay}
\email{kraszna@atomki.hu}
\author{M. Csatl\'os}
%\author{L. Csige}
\author{J. Guly\'as}
%\author{A.~Krasznahorkay}
%\altaffiliation{Currently working at CERN, Geneva, Switzerland}
\author{B. Kert\'esz}
\author{A.~Krasznahorkay}
\altaffiliation{Currently working at CERN, Geneva, Switzerland}
\author{J. Moln\'ar}
\author{I. Rajta}
\author{J. Tim\'ar}
\author{I. Vajda}
\affiliation{Institute of Nuclear Research  (ATOMKI),
  P.O. Box 51, H-4001 Debrecen, Hungary}
\author{M.N. Harakeh}
\affiliation{Nuclear Energy Group, ESRIG, University of Groningen,
Zernikelaan 25, 9747 AA Groningen, The Netherlands}

%%%
\begin{abstract}

Angular correlation spectra of $e^+e^-$ pairs produced
in the $^{7}$Li($p$,$\gamma$)$^{8}$Be nuclear reaction have been studied
at the sharp $E_p$= 441 keV resonance as well as at $E_p$= 650 keV, 800
keV and 1100 keV proton beam energies.  The spectra measured at the
resonance can be understood through the M1 internal
pair creation process, but in the case of the off-resonance regions
(direct proton capture) significant anomalies were observed in the
$e^+e^-$ angular correlations supporting the X17 hypothetical particle
creation and decay.
%\PACS{21.10.Re \and 24.50.+g \and 25.45.Hi  \and 25.85.Ge \and 27.90.+b} 
\end{abstract}

\maketitle
%%%

\section{Introduction}
Recently, we studied electron-positron angular correlations for the
17.6~MeV and 18.15~MeV transitions in $^8$Be and an anomalous angular
correlation was observed for the 18.15~MeV transition \cite{kr16}.
This was interpreted as the creation and decay of an intermediate
bosonic particle with a mass of
$m_\mathrm{X}c^2$=16.70$\pm$0.35(stat)$\pm$0.5(sys)~MeV, which is now
called X17. The possible relation of the X17 boson to the dark-matter
problem triggered an enormous interest in the wider physics community
\cite{da19,ins}.

The first theoretical interpretation of the experimental results was
provided by Feng et al. \cite{fe16,fe17}.  They generalized the
theory of the dark photon so that the new particle, which in the
literature was named X17, could be coupled not only to the electric
charge but also to the quarks. Coupling constants were determined
using our, and previously obtained, experimental data. They called
their theory protophobic because the X17 boson was coupled much weaker 
to protons than to neutrons.  
They predicted that the X17 particle should also be created in
the 17.6 MeV transition of the $^8$Be nucleus, with a branching ratio
2.3 times smaller than in the case of the 18.15 MeV
transition.  However, in our original publication we did not find any
anomaly in that transition, which could confirm this prediction.

In 2017, we re-investigated the $^8$Be anomaly with an improved, and
independent setup, and confirmed the signal of the assumed X17
particle \cite{kra18,kra19}. We studied also the 17.6 MeV M1
transition to check the different theoretical predictions, but
obtained different branching ratios \cite{kr137,kr141,kr17}. 

Recently, we also observed a similar anomaly in $^4$He
\cite{kra19,fi20,kr20,kr21}.  The signal could be described by the
creation and subsequent decay of a light particle during the proton
capture process on $^3$H to the ground state of the $^{4}$He
nucleus. The derived mass of the particle ($m_\mathrm{X}c^2 = 16.94
\pm 0.12$(stat.)$\pm 0.21$(syst.)~MeV) agrees well with that of the proposed
X17 particle.  It was also shown that the branching ratios of
the X17 particle are identical within uncertainties for  three
beam energies, proving that the X17 particle was most likely formed in
direct proton capture, which has a dominant multipolarity of E1.
Our results obtained
for $^4$He at different beam energies agree well with the present 
theoretical results calculated with ab-initio models by Viviani et al.
\cite{vi21}.

Zhang and Miller \cite{zh21} studied the protophobic
vector boson explanation in $^8$Be, by deriving an isospin relation between the
coupling of photon and X17 to nucleons. They concluded that the X17
production may be dominated by direct-capture (E1) transitions and a smooth
energy dependence is predicted for all proton beam energies above
the 17.6~MeV  J$^\pi$=1$^+$ resonance \cite{zh21}. (However, for the
resonance they found the M1-induced X17 production is also very important.)

The aim of the present work was to study the off-resonance region for
the  $^{7}$Li($p$,$e^+e^-$)$^{8}$Be reaction
in order to check if the X17 particle is created at these energies as well,
as predicted by Zhang and Miller
 \cite{zh21}.

%%%%

\section{Experiments}

A proton beam with a typical current of 5 $\mu$A bombarded LiF and
Li$_2$O targets for about 50 hours for each bombarding energy.  The
target thickness for the on-resonance measurement was 30 $\mu g/cm^2$,
and for the off-resonance measurements 300 $\mu g/cm^2$. All of the
targets were evaporated onto aluminum strips with thicknesses of 10
$\mu$m.

\begin{figure}
\resizebox{0.45\textwidth}{!}{
  \includegraphics{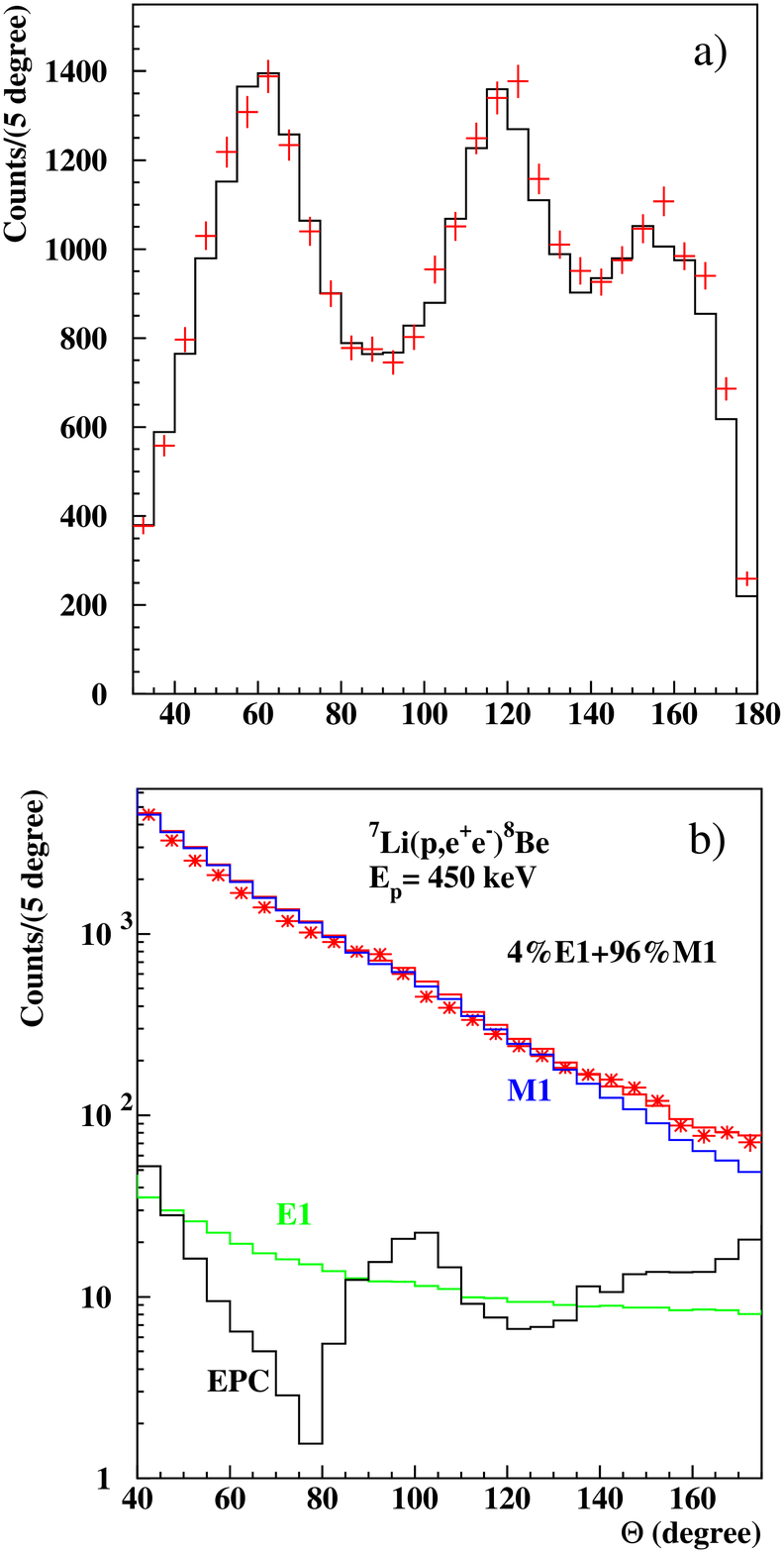}}
  \caption{\it a: Detector response for the setup as a function of
    correlation angle ($\Theta$) for isotropic emission of e$^+$e$^-$
    pairs (red crosses) compared with the results of the Monte
    Carlo simulations (black-line histogram) as explained in the
    text. b: $e^+e^-$ angular correlations obtained for the 17.6~MeV
    transition of $^8$Be by using thin target backing compared to the
    simulations performed for E1 and M1 IPC, as well as for the EPC created
    by the $\gamma$-rays on the different materials around the target.}
  \label{fig:lif-calibration1}
\end{figure}

The Plexiglas rods, we used previously for holding the targets
\cite{kr16,gu16}, were replaced with Al rods to get a better cooling
of the targets to prevent the diffusion of Li/Li$_2$O into the target
backing.  In this way we managed to significantly increase the
lifetime of the targets, which made it possible to perform on-
and off-resonance measurements with the same target. Unfortunately, such Al
rods produced a somewhat larger background induced by the
$\gamma$-rays on the rods, as shown in Fig.~\ref{fig:lif-calibration1}
b), than what we observed in Fig. 9 of Ref. \cite{gu16}.

Our previous detector setup \cite{kr16,gu16} has recently been
upgraded. The details of the upgrade are described in detail in
\cite{kr21}.  
Time and energy signals of the scintillators, as well as the time,
energy and position signals of the DSSD detectors were recorded.

In order to search for the assumed X17 particle, both the energy-sum
spectrum of the $e^+e^-$ pairs measured by the telescopes, and their
angular correlations, determined by the DSSD detectors, have been
analyzed. For the real ``signal'' events we always required that the
energy-sum for the $e^+e^-$-pairs should be equal to the transition
energy, which we want to investigate.

The energy calibration of the telescopes, the energy and position
calibrations of the DSSD detectors, the Monte Carlo simulations as
well as the acceptance calibration of the whole $e^+e^-$ coincidence
pair spectrometer were performed in a similar way as we described in
Ref. \cite{kr21}.

Reasonably good agreement was obtained with the results of the Monte
Carlo simulations, as presented in Fig.~\ref{fig:lif-calibration1} for
the present setup.  The average difference is within $\approx 3.0$\%
in the 40$^\circ$~-~170$^\circ$ range.

In order to test the accuracy of these %-- naturally not perfect --
simulations for describing our experiments, we made measurements with
the $^{7}$Li($p$,$\gamma$)$^{8}$Be reaction.  The results for the
angular correlations from this data obtained at the E$_p$~=~441~keV
resonance (red dots with error bars) are shown in
Fig.~\ref{fig:lif-calibration1} b), together with the corresponding
IPC Monte Carlo simulation (histogram) coming mostly from the M1
nuclear transition. The contribution coming from the External Pair
Creation (EPC) of the 17.6 MeV $\gamma$-rays is shown by solid
black-line histogram.  We note here that the direct-capture
contribution is negligible compared to the M1 IPC due to the large
resonance capture cross section and the thin target.  The ratio of the
event numbers used for the simulations are determined by the internal
pair creation coefficient of the 17.6 MeV M1 transition.

As it can be seen in Fig.~\ref{fig:lif-calibration1}, the simulation
of this single (IPC) process manages to describe the shape of the
e$^+$e$^-$ angular correlation data distribution accurately, and the
contribution of EPC created on the different parts of the spectrometer
is reasonably low. It is especially true for the Al backing of the
target.  We performed simulations without such backing and with
backing. Their difference was smaller than 10\% of the full EPC
contribution in the $40^\circ \leq [Q] \leq 175^\circ$ angular range.

\section{Experimental results}

At the low bombarding energies we used, the monitoring $\gamma$-ray
spectra observed in the 13.5 MeV - 20 MeV energy range were very clean
as shown in Fig. 2 a).  The corresponding energy-sum spectra of the
$e^+e^-$ pairs measured by the telescopes are shown in Fig. 2 b). The
``signal'' region for E(sum) was chosen to be very wide, from 13.5 MeV
to 20.0 MeV including both the transition to the ground state and the
transition to the first-excited state of $^8$Be.

\begin{figure}
\resizebox{0.4\textwidth}{!}{%
  \includegraphics{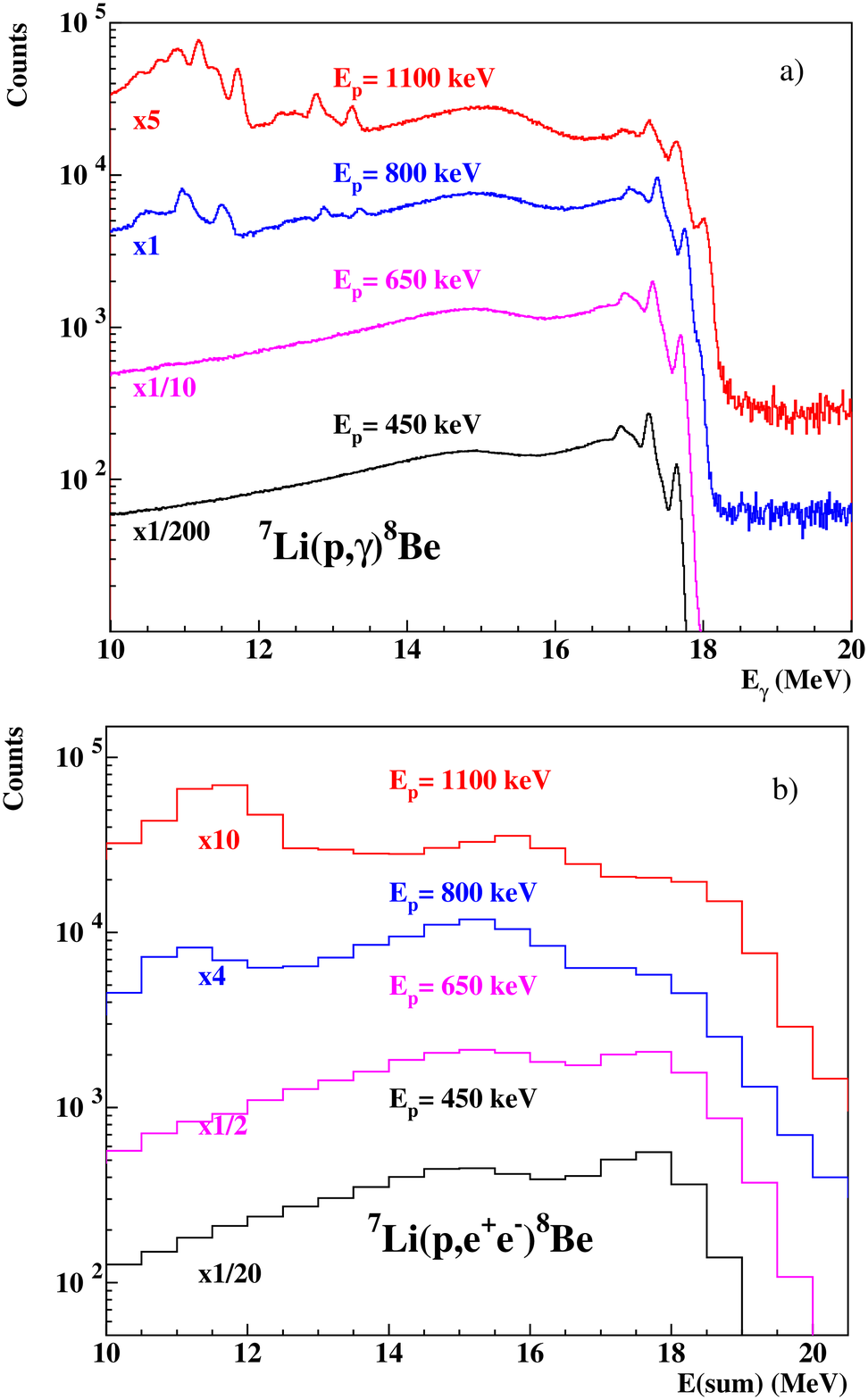}}         
\caption{Typical $\gamma$-ray energy spectra (a) and energy-sum
  spectra of the $e^+e^-$ pairs derived, respectively, for 450~keV,
  650~keV, 800~keV and 1100 keV bombarding energies}
 \label{fig:gamma}
\end{figure}

The angular correlations of the $e^+e^-$ pairs were determined from
the position data of the DSSD detectors for each beam energy.  The
Cosmic Ray Background CRB contributions were subtracted. Considering
the kinematics of the $e^+e^-$ pair-creation process, we also required
the following condition for the asymmetry parameter: $-0.3\leq
y=(E_{e^+} - E_{e^-})/(E_{e^+} + E_{e^-}) \leq 0.3$, where $E_{e^+}$
and $E_{e^-}$ denote the kinetic energies of the positron and
electron, respectively. The raw angular correlations were then
corrected for the detector response shown in
Fig.~\ref{fig:lif-calibration1}.

 \begin{figure}
  \resizebox{0.5\textwidth}{!}{% 
\includegraphics{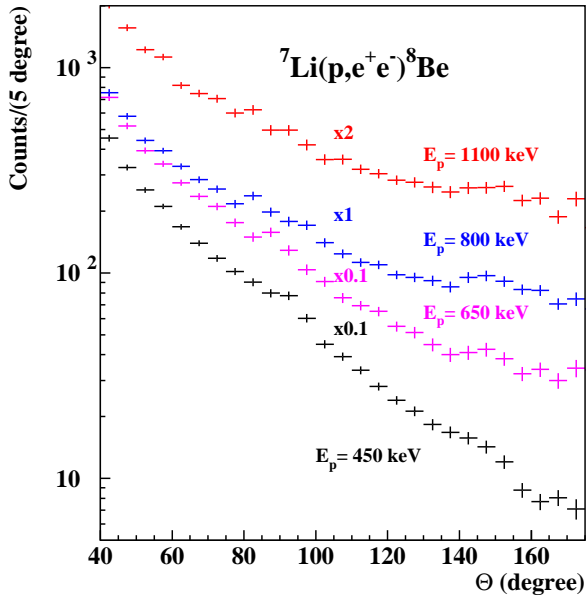}}
\caption{
Angular correlations of the $e^+e^-$ pairs for the ``Signal" region.
Symbols with error bars indicate experimental data
measured in the $^{7}$Li($p,\gamma$)$^{8}$Be reaction at different
proton beam energies.}
\label{raw-ang-distr}
\end{figure}

The resulting angular-correlation spectra are indicated in
Fig.~\ref{raw-ang-distr} by dots with error bars for E$_p$= 450 keV
(black) and 650 keV (magenta), 800 keV (blue) and 1100 keV (red). For
better visibility, some spectra are 
multiplied by factors as shown in the figure.

The slopes of the angular correlations differ significantly. They are
expected to have contributions from M1 multipolarity transitions
coming from the resonant proton capture (17.6 and 18.15 MeV $J^\pi =
1^+$ states) as well as from the E1 multipolarity transitions
resulting from the direct proton-capture process.  Since the simulated
angular correlation drops much steeper for the M1 than for the E1
multipolarities as shown in Fig.~\ref{fig:lif-calibration1}, the
experimental angular correlation could be fitted by the linear
combination of the two (M1 and E1) simulated curves in the 40-130
degree angular range, where no anomaly is expected in
Fig.~\ref{raw-ang-distr}.  The results of the fits are shown in
Fig.~\ref{background}., which can be reasonably well explained as
follows:

\begin{figure}
  \resizebox{0.5\textwidth}{!}{%
  \includegraphics{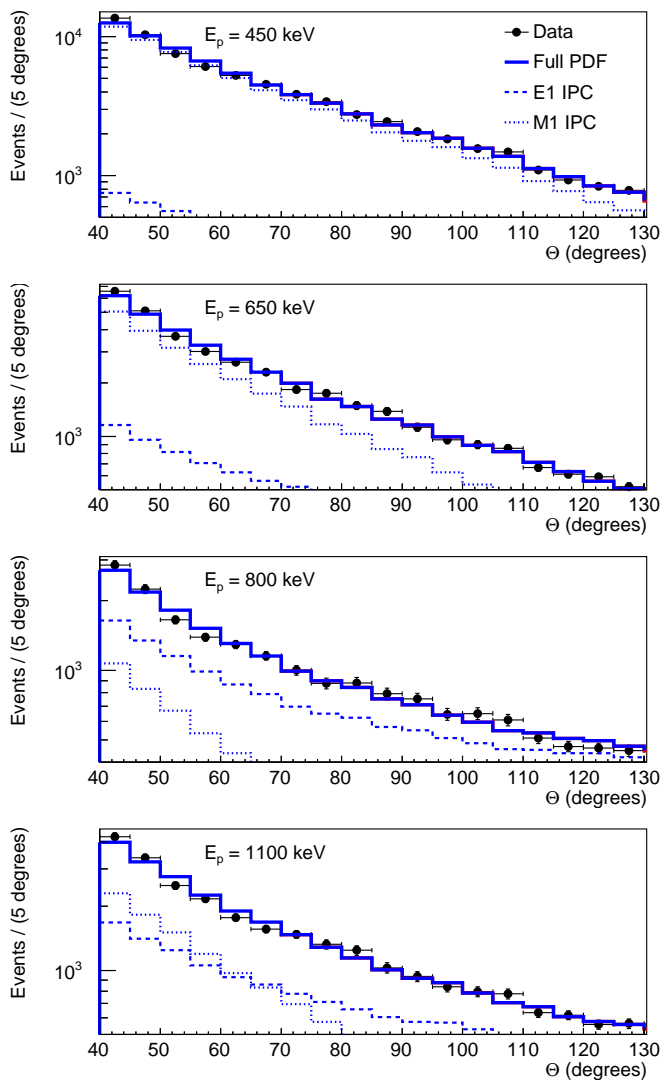}}\hspace{0.5cm}
    \caption{
Angular correlations of the $e^+e^-$ pairs for the ``Signal" region.
Symbols with error bars indicate experimental data
measured in the $^{7}$Li($p,\gamma$)$^{8}$Be reaction at different
proton beam energies, while solid-line histograms correspond to the
respective data obtained in the simulations described in the text.}
\label{background}
\end{figure}

\begin{enumerate}
\item the M1 contribution is the largest for the $1^+$ 441 keV (E$_p$= 450 keV)
resonance and dominates the distribution. 

\item The E1 distribution resulting
from the proton direct capture starts to dominate at E$_p$=650 keV, but
since the proton energy loss in the target is about 200 keV, the M1
distribution from the strong 441 keV resonance is still visible.

\item The contribution of the E1 distribution is the largest at E$_p$=800
keV. There is only 9.5\% M1 contribution, which may be coming from the tail
of the 1030 keV resonance.  

\item The contribution of the M1 distribution
increased by a factor of more than 3 at E$_p$=1100 keV, when we are
on the top of the 1030 keV $1^+$ resonance.  
\end{enumerate}

The fitted E1 and M1 simulated angular correlations are assumed to be
valid also at larger angles up to 175 degrees (the last bin of the
distributions where the acceptance of the spectrometer is the smallest
was rejected from the fit) and the anomaly is defined as the
difference of the experimental and simulated distributions.

\begin{figure}
  \resizebox{0.5\textwidth}{!}{%
\includegraphics{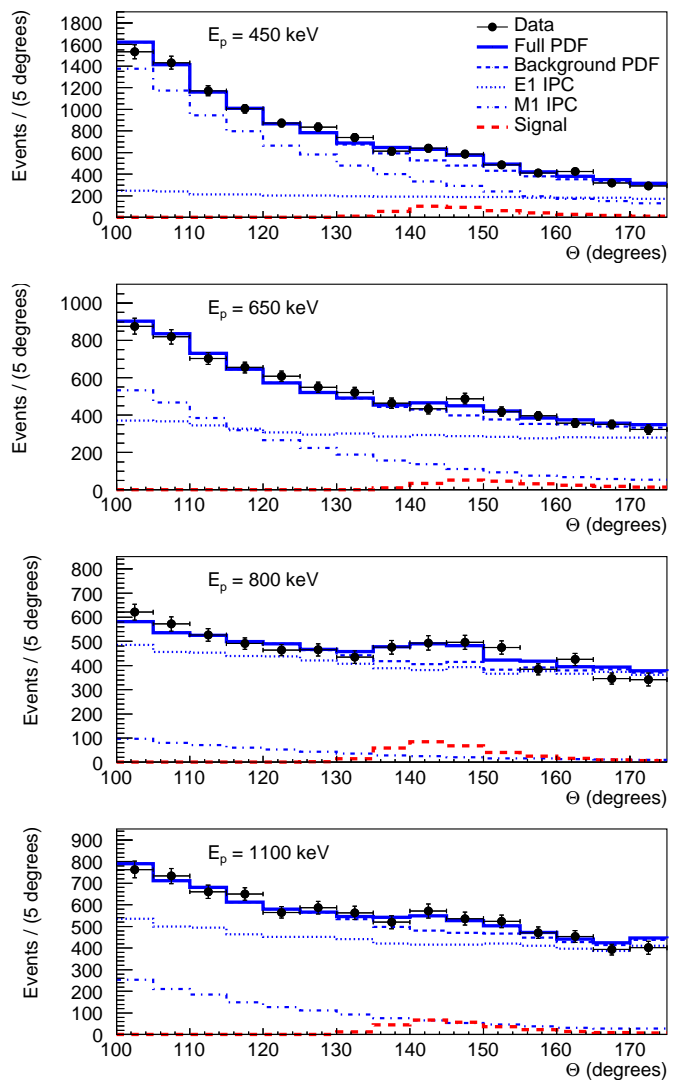}}          
  \caption{See the figure caption of Fig. 4. The fit was performed for the
    angular region of the anomaly.}
\label{sig-bestfits}
\end{figure}

The anomaly around 140$^\circ$ was then fitted with the mixed E1+M1
angular correlation with the mixing ratio obtained for the
40$^\circ$-175$^\circ$ angular range, and the expected $e^+e^-$ decay
of the assumed X17 particle.  The fit was performed with RooFit
\cite{Verkerke:2003ir} in a similar way as we did in our previous work
\cite{kr21}.  The results of the fits corresponding to the different
proton bombarding energies are shown in Fig.~\ref{sig-bestfits}.

The fitting parameters for the E1 and M1 and IPC distributions
as well as the contribution of the X17 particle are summarized in
Table~\ref{results}. The significance of the fits on average is:
4.5$\sigma$.

\begin{table}[h]
  \centering
  \caption{The fitted mass [m(X17)] and the integrated yields I(X17),
    I(E1) and I(M1) of the X17 and the E1 and the M1 contributions. The
    ratio of I(X17)/I(E1) is also listed [B(X17)].}
  \label{results}
  \begin{tabular}{llllll}
    \hline\hline E$_p$ & m(X17) &I(X17) & I(E1) & I(M1) & B(X17)
    \\ (keV) &(MeV/c$^2$) & & & & \\ \hline 450 & 16.6(3) & 43(49) &
    30(25) & 79(2) & 1.4(16) \\ 650 & 16.94(14) & 24(16) & 46(5) &
    32(4) & 0.5(3) \\ 800 & 16.81(9) & 33(10) & 62(4) & 5.9(4) &
    0.53(14) \\ 1100 & 17.11(12) & 28(8) & 66(2) & 15(1) & 0.41(13)
    \\ \hline\hline
  \end{tabular}
\end{table}

Checking the fitted parameters, the ratio of I(M1)/I(E1) determined at
the E$_p$= 1030 keV resonance compared to the one determined at 800
keV is about a factor of two larger, as suggested recently by Hayes et
al. in Fig. 3. of Ref. \cite{ha21}.  It seems also that the amplitude
of the X17 contribution, I(X17), correlates with the amplitude of the
E1 multipolarity, I(E1), and not with the M1 one as suggested in
Ref. 1.  This result supports the vector character of the X17 particle
and not the axial-vector one as suggested in earlier publications.

Table~\ref{results} displays only the statistical errors.
The systematic uncertainties were estimated from the
simulations in a similar way to the previous work \cite{kr21}, and
obtained: $\Delta m_\mathrm{X}c^2$(syst.)~$= \pm 0.2$~MeV.

\section{Summary}

We have studied the energy-sum and angular correlation spectra of
$e^+e^-$ pairs produced in the $^7$Li($p$,$e^+e^-$)$^8$Be reaction at
E$_p$= 450, 650, 800 and 1100~keV proton energies.  The main features
of the spectra can be understood rather well by taking into account
the internal pair creations following the M1 radiations coming from
the decay of the $1^+$ states and the E1 ones coming from the direct
proton capture on the target.

 We observed a peak-like anomalous excess of $e^+e^-$ pairs in the
 angular correlation spectra around 140$^\circ$ at each beam
 energy. This $e^+e^-$ excess can be described by the creation and
 subsequent decay of a light particle, created during the
 proton-capture process to the ground state of the $^{8}$Be
 nucleus. The derived mass of the particle ($m_\mathrm{X}c^2 = 16.95
 \pm 0.10$(stat.)$\pm 0.21$(syst.)~MeV) agrees well with that of the
 X17 particle, which we recently suggested \cite{kr16,kra18,kra19} for
 describing the anomaly observed in $^8$Be.

 The contribution anomalous excess at 450 keV was much smaller than in
 the other measurements, which seems to contradict with the
 theoretical prediction of Zhang and Miller \cite{zh21}.

It seems that the particle is created in the direct proton-capture
process and not in the M1 decay of the 17.6 and 18.15 MeV J$^\pi =
1^+$ states.  Our present results obtained for $^8$Be at different
beam energies above E$_p$=450~keV agree well with the prediction of
Zhang and Miller \cite{zh21} and do not invalidate the protophobic
vector boson interpretation of Feng et al.  \cite{fe16}.

%With the X17 particle appearing to be created in the E1 transition, we
%are forming plans to search for it in the decay of Giant Dipole
%Resonance (GDR) excitations of different nuclei as well.

\section{Acknowledgements}
We wish to thank Z. Pintye for the mechanical design of the
experiment.  This work has been supported by the Hungarian NKFI
Foundation No.\, K124810 and by the GINOP-2.3.3-15-2016-00034 and

\noindent GINOP-2.3.3-15-2016-00005 grants.

\section{Appendix: Comparison with our previous results}

The difference between our previous results \cite{kr16} and the
present ones needs some explanation. Especially, the angular
correlation spectrum at E$_p$=800~keV seems to differ considerably
from the one reported from our previous experiment \cite{kr16}. We did
not report about an anomaly at this energy. The contradiction can be
resolved taking into account that the observation or non-observation
of the anomaly depends strongly on how one models the background.

 The shape of the background
depends strongly on the ratio of the induced M1 and E1 IPC processes
. It turned out that properties of the used targets can
considerably modify this ratio, which will be discussed further below.

\begin{figure}
\resizebox{0.5\textwidth}{!}{%
        \includegraphics{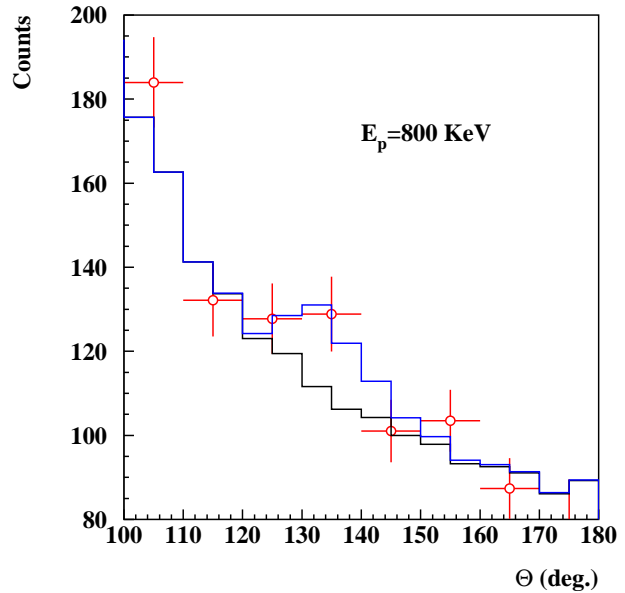}}          
\caption{\it
  Experimental angular correlation of the $e^+e^-$ pairs obtained  from the
  $^{7}$Li($p,\gamma$)$^{8}$Be reaction at E$_p$ = 800~keV 
  proton beam energy  \cite{kr16} analyzed with a new background explained in
  the text.}

 \label{comp}
\end{figure}

The results of our previous angular-correlation measurement
\cite{kr16} performed at E$_p$=800 keV are shown in Fig.~\ref{comp}.
Red circles with error bars indicate the previous results, the black
histogram represents the fitted background using the E1/M1 ratio
observed in the present measurement, while the blue histogram
represent the results of the simulations performed for the X17 boson.

It can be clearly seen in Fig.~\ref{comp} that with the new
background, the peak-like anomaly is also visible at E$_p$=800~keV in
our previous measurement, not only at E$_p$=1100~keV as we wrote in
our previous publication. In this way, the two data sets do not
actually contradict each other.

In Fig. 3 of Ref. \cite{kr16}, measurements at E$_p$ = 1040 keV and
E$_p$ = 1100 keV were performed first. For these measurements, metal
Li evaporated on Al backing was used as the target.
After the targets were made, and 
until they were used, they could come into contact with air several
times and thus become heavily oxidized, as indicated by their milky
white color.

After the anomaly was observed at these energies, the off-resonance
measurements were performed with thicker targets, and the
transportation was made in an Argon atmosphere with more care taken so
as not to oxidize the targets. We used thicker targets to offset the
much smaller cross sections off resonance. However, by heating the
target through bombardment by the proton beam, the metal Li diffused
into the Al backing.  Hence, the actual target thickness became equal
to the backing thickness, which was 10 $\mu$m, and in which the 800
keV protons were already completely stopped. Accordingly, we obtained
significant amounts of $e^+ e^-$ pairs not only from the off-resonance
region, but also from the strong 441 keV resonance, whose
multipolarity being M1, increased the background at small
angles. (Examining the gamma spectra measured at that time, the 17.6
MeV transition from the 441 keV resonance can also be intensively
observed.) This was the reason that the shape of the $e^+ e^-$
background was assumed to be similar during the on- and off-resonance
measurements in Ref. \cite{kr16}, although it should have been
different as the M1 content should have been smaller at E$_p$ = 800
keV than at E$_p$ = 1100 keV. By using the present results for the
shape of the background the anomaly becomes visible also at E$_p$ =
800 keV, like in the present data. However, it should be mentioned
that the new background underestimates the background of Ref.
\cite{kr16} at smaller angles, which was coming mostly from the 17.6
MeV M1 transition in that case.

%%%%%%%%%%%%%%%%%%%%%%%%%%%%%%%%%%%%%%%%%%%%%%%%%%%%%%%%%%%%%%%%%%%%%%%

%\section*{References}

\end{document}